\documentclass[conference]{IEEEtran}

\IEEEoverridecommandlockouts
\DeclareRobustCommand{\IEEEauthorrefmark}[1]{\smash{\textsuperscript{\footnotesize #1}}}
\usepackage{enumitem} % Load enumitem package

\usepackage{cite}
\usepackage{amsmath,amssymb,amsfonts}
\usepackage{algorithmic}
\usepackage{graphicx, subcaption}
\usepackage{siunitx}
\usepackage{textcomp}
\RequirePackage[hyphens]{url}
\usepackage{xcolor}
\usepackage[colorinlistoftodos,prependcaption,textsize=tiny]{todonotes}
\usepackage[normalem]{ulem}
\usepackage{tikz}
\usetikzlibrary{arrows,arrows.meta,decorations.pathmorphing,backgrounds,positioning,fit,petri,calc,shapes.geometric,scopes,fit,shapes.geometric,math}

\usepackage[hidelinks,colorlinks=false,pdfborder={0 0 0}]{hyperref}
\usepackage{orcidlink}
\usepackage{setspace}

\usepackage[toc=false,indexonlyfirst=true,acronyms]{glossaries}
\newacronym{ml}{ML}{machine learning}
\newacronym{jepa}{JEPA}{joint-embedding predictive architecture}
\newacronym{vit}{ViT}{vision transformer}
\newacronym{vae}{VAE}{variational autoencoder}
\newacronym{mae}{MAE}{masked autoencoder}
\newacronym{er}{ER}{experience replay}
\newacronym{lts}{LTS}{long-term storage}
\newacronym{lwfa}{LWFA}{laser wakefield acceleration}
\newacronym{pic}{PIC}{particle-in-cell}
\newacronym{ddp}{DDP}{distributed data parallel}
\newacronym{gcd}{GCD}{graphics compute die}
\newacronym{gil}{GIL}{global interpreter lock}
\newacronym{pca}{PCA}{principal component analysis}
\newacronym{pc}{PC}{principal component}
\newacronym{rdma}{RDMA}{remote direct memory access}
\newacronym{rope}{RoPE}{rotary positional embedding}
\newacronym{ssl}{SSL}{self-supervised learning}
\newacronym{sst}{SST}{Sustainable Staging Transport}
\newacronym{stii}{STII}{self-truncated ionization injection}

\newcommand\jk[1]{\textcolor{green!70!black}{#1}}
\newcommand\jksout{\bgroup\markoverwith{\textcolor{green!70!black}{\rule[0.5ex]{2pt}{0.4pt}}}\ULon}

\def\atfranz#1{\textcolor{blue}{@FP: #1}}
\def\atnikhil#1{\textcolor{green!70!black}{@NR: #1}}

\newcommand\richardsout{\bgroup\markoverwith{\textcolor{purple}{\rule[0.5ex]{2pt}{0.4pt}}}\ULon}
\newcommand\richardsUL{\bgroup\markoverwith{\textcolor{purple}{\rule[-0.2ex]{2pt}{0.4pt}}}\ULon}

\def\BibTeX{{\rm B\kern-.05em{\sc i\kern-.025em b}\kern-.08em
    T\kern-.1667em\lower.7ex\hbox{E}\kern-.125emX}}

\begin{document}

% Enter the title of your submission. Please use standard title capitalization. Use initial-caps for most words except prepositions. Also use initial-caps for both parts of hyphenated words. 
\title{The Recycler: In-Transit Machine Learning on High-Fidelity Data from
Simulation Campaigns as an Optimized Approach for Building Frontier Models}

%\iffalse
\author{
\IEEEauthorblockN{Nikhil Rao} 
\IEEEauthorblockA{\textit{University of Delaware} \\
Newark, DE,  United States of America \\
ORCID: 0000-0002-3560-9428} % nikhilr@udel.edu
\and
\IEEEauthorblockN{Mark Berrill} \\
\IEEEauthorblockA{\textit{Oak Ridge National Laboratory}\\
Oak Ridge, TN, United States of America \\
ORCID: 0000-0002-4525-3939} % BERRILLMA@ORNL.GOV
\and
\IEEEauthorblockN{Ankush Checkervarty}
\IEEEauthorblockA{\textit{Helmholtz-Zentrum Dresden-Rossendorf (HZDR)} \\
Dresden, Germany \\
ORCID: 0009-0004-4653-0004} % a.checkrvarty@hzdr.de
\and
\IEEEauthorblockN{Alexander Debus}
\IEEEauthorblockA{\textit{Helmholtz-Zentrum Dresden-Rossendorf (HZDR)} \\
Dresden, Germany \\
ORCID: 0000-0002-3844-3697} % a.debus@hzdr.de
\and
\IEEEauthorblockN{Greg Eisenhauer}
\IEEEauthorblockA{\textit{Georgia Institute of Technology} \\
Atlanta, GA, United States of America  \\
ORCID: 0000-0002-2070-043X} % eisen@cc.gatech.edu
\and
\IEEEauthorblockN{Ana Gainaru}
\IEEEauthorblockA{\textit{LANL}\\
Los Alamos, NM,  United States of America \\
ORCID: 0000-0002-1375-9468} % againaru@lanl.gov
\and
\IEEEauthorblockN{Scott Klasky} 
\IEEEauthorblockA{\textit{Oak Ridge National Laboratory}\\
Oak Ridge, TN, United States of America \\
ORCID: 0000-0003-3559-5772} % klasky@ornl.gov
\and 
\IEEEauthorblockN{Julian Lenz}
\IEEEauthorblockA{\textit{Center for Advance Systems Understanding (CASUS)} \\
Görlitz, Germany \\
ORCID: 0000-0001-5250-0005} % j.lenz@hzdr.de
\and
\IEEEauthorblockN{Tapish Narwal}
\IEEEauthorblockA{\textit{Center for Advance Systems Understanding (CASUS)} \\
Görlitz, Germany \\
ORCID: 0000-0002-5726-2330} % t.narwal@hzdr.de 
\and
\IEEEauthorblockN{Vedhas Pandit}
\IEEEauthorblockA{\textit{Helmholtz-Zentrum Dresden-Rossendorf (HZDR)} \\
Dresden, Germany \\
ORCID: 0000-0002-1983-8140} % v.pandit@hzdr.de
\and
\IEEEauthorblockN{Richard Pausch}
\IEEEauthorblockA{\textit{Helmholtz-Zentrum Dresden-Rossendorf (HZDR)} \\
Dresden, Germany \\
ORCID: 0000-0001-7990-9564} % r.pausch@hzdr.de
\and
\IEEEauthorblockN{Norbert Podhorszki}
\IEEEauthorblockA{\textit{Oak Ridge National Laboratory} \\
Oak Ridge, TN, United States of America \\
ORCID: 0000-0001-9647-542X} % pnb@ornl.gov
\and
\IEEEauthorblockN{Franz Pöschel}
\IEEEauthorblockA{\textit{Center for Advance Systems Understanding (CASUS)} \\
Görlitz, Germany \\
ORCID: 0000-0001-7042-5088} % f.poeschel@hzdr.de
\and
\IEEEauthorblockN{Ulrich Schramm}
\IEEEauthorblockA{\textit{Helmholtz-Zentrum Dresden-Rossendorf (HZDR)} \\
Dresden, Germany \\
ORCID: 0000-0003-0390-7671} % u.schramm@hzdr.de
\and
\IEEEauthorblockN{Klaus Steiniger}
\IEEEauthorblockA{\textit{Helmholtz-Zentrum Dresden-Rossendorf (HZDR)} \\ 
\textit{Center for Advance Systems Understanding (CASUS)} \\
Dresden, Germany \\
ORCID: 0000-0001-8965-1149} % k.steiniger@hzdr.de
\and
\IEEEauthorblockN{Jessica Tiebel}
\IEEEauthorblockA{\textit{Helmholtz-Zentrum Dresden-Rossendorf (HZDR)} \\ 
Dresden, Germany \\
ORCID: 0000-0001-9235-5287} % j.tiebel@hzdr.de
\and
\IEEEauthorblockN{René Widera}
\IEEEauthorblockA{\textit{Helmholtz-Zentrum Dresden-Rossendorf (HZDR)} \\ 
Dresden, Germany \\
ORCID: 0000-0003-1642-0459} % r.widera@hzdr.de
\and
\IEEEauthorblockN{Nico Wrobel}
\IEEEauthorblockA{\textit{Helmholtz-Zentrum Dresden-Rossendorf (HZDR)} \\ 
\textit{Center for Advance Systems Understanding (CASUS)} \\
Dresden, Germany \\
ORCID: 0009-0008-7039-6377} % n.wrobel@hzdr.de
\and
\IEEEauthorblockN{Yongqi Xiang}
\IEEEauthorblockA{\textit{Helmholtz-Zentrum Dresden-Rossendorf (HZDR)} \\
Dresden, Germany \\
ORCID: 0009-0005-5578-2067} % y.xiang@hzdr.de
\and
\IEEEauthorblockN{Michael Bussmann}
\IEEEauthorblockA{\textit{Helmholtz-Zentrum Dresden-Rossendorf (HZDR)} \\ 
\textit{Center for Advance Systems Understanding (CASUS)} \\
Dresden, Germany \\
ORCID: 0000-0002-8258-3881} % m.bussmann@hzdr.de
\and
\IEEEauthorblockN{Sunita Chandrasekaran} 
\IEEEauthorblockA{\textit{University of Delaware} \\
Newark, DE,  United States of America \\
ORCID: 0000-0002-3560-9428} % schandra@udel.edu
\and
\IEEEauthorblockN{Jeffrey Kelling}
\IEEEauthorblockA{\textit{Helmholtz-Zentrum Dresden-Rossendorf (HZDR)} \\
\textit{Chemnitz University of Technology} \\
Dresden, Germany \\
ORCID: 0000-0003-1761-2591} % j.kelling@hzdr.de
}

\author{
\IEEEauthorblockN{
Nikhil Rao\IEEEauthorrefmark{1}, %\orcidlink{0000-0003-2639-9892} % nikhilr@udel.edu
Mark Berrill\IEEEauthorrefmark{2}, %\orcidlink{0000-0002-4525-3939} % % BERRILLMA@ORNL.GOV 
Ankush Checkervarty\IEEEauthorrefmark{3}, %\orcidlink{0009-0004-4653-0004} % a.checkrvarty@hzdr.de
Alexander Debus\IEEEauthorrefmark{3}, %\orcidlink{0000-0002-3844-3697} % a.debus@hzdr.de
\\
Greg Eisenhauer\IEEEauthorrefmark{4}, %\orcidlink{0000-0002-2070-043X} % eisen@cc.gatech.edu
Ana Gainaru\IEEEauthorrefmark{5}, %\orcidlink{0000-0002-1375-9468} % againaru@lanl.gov 
Scott Klasky\IEEEauthorrefmark{2}, %\orcidlink{0000-0003-3559-5772} % klasky@ornl.gov
Julian Lenz\IEEEauthorrefmark{3,6}, %\orcidlink{0000-0001-5250-0005} % j.lenz@hzdr.de
Tapish Narwal\IEEEauthorrefmark{3,6}, %\orcidlink{0000-0002-5726-2330} % t.narwal@hzdr.de 
Vedhas Pandit\IEEEauthorrefmark{3}, %\orcidlink{0000-0002-1983-8140} % v.pandit@hzdr.de
Richard Pausch\IEEEauthorrefmark{3}, %\orcidlink{0000-0001-7990-9564} % r.pausch@hzdr.de
\\
Norbert Podhorszki\IEEEauthorrefmark{2}, %\orcidlink{0000-0001-9647-542X} % pnb@ornl.gov
Franz Pöschel\IEEEauthorrefmark{3,6}, %\orcidlink{0000-0001-7042-5088} % f.poeschel@hzdr.de
Ulrich Schramm\IEEEauthorrefmark{3,7}, %\orcidlink{0000-0003-0390-7671} % u.schramm@hzdr.de
Klaus Steiniger\IEEEauthorrefmark{3,6}, %\orcidlink{0000-0001-8965-1149} % k.steiniger@hzdr.de
Jessica Tiebel\IEEEauthorrefmark{3,7}, %\orcidlink{0000-0001-9235-5287} % j.tiebel@hzdr.de
René Widera\IEEEauthorrefmark{3}, %\orcidlink{0000-0003-1642-0459} % r.widera@hzdr.de
\\
Nico Wrobel\IEEEauthorrefmark{3,6}, %\orcidlink{0009-0008-7039-6377} % n.wrobel@hzdr.de
Yongqi Xiang\IEEEauthorrefmark{3}, %\orcidlink{0009-0005-5578-2067} % y.xiang@hzdr.de
Michael Bussmann\IEEEauthorrefmark{3,6}, %\orcidlink{0000-0002-8258-3881} % m.bussmann@hzdr.de
Sunita Chandrasekaran\IEEEauthorrefmark{1}, %\orcidlink{0000-0002-3560-9428} % schandra@udel.edu
Jeffrey Kelling\IEEEauthorrefmark{3,8} %\orcidlink{0000-0003-1761-2591} % j.kelling@hzdr.de
}
\\
    \IEEEauthorblockA{
    \IEEEauthorrefmark{1}\textit{University of Delaware} Newark, DE,  United States of America\\
\IEEEauthorrefmark{2}\textit{Oak Ridge National Laboratory} Oak Ridge, TN, United States of America\\
\IEEEauthorrefmark{3}\textit{Helmholtz-Zentrum Dresden-Rossendorf (HZDR)}, Dresden, Germany\\
\IEEEauthorrefmark{4}\textit{Georgia Institute of Technology}, Atlanta, GA, United States of America\\
\IEEEauthorrefmark{5}\textit{Los Alamos National Laboratory} Los Alamos, NM,  United States of America\\
\IEEEauthorrefmark{6}\textit{Center for Advance Systems Understanding (CASUS)}, Görlitz, Germany\\
\IEEEauthorrefmark{7}\textit{Dresden University of Technology (TUD)}, Dresden, Germany\\
\IEEEauthorrefmark{8}\textit{Chemnitz University of Technology}, Chemnitz, Germany
}
}

\newcommand\sectionGuide[3]{
\iffalse
 \textit{#1} % volume limit, e.g. "1 paragraph = 0.5 column"
 \textbf{#2} % assignee(s)
 \\
 {\color{gray}
 #3 % content/guidelines
 }
\fi
}

\maketitle

\begin{abstract}
We report on the first realization of in-transit machine learning using data streamed from multiple simulations running in parallel to the training on the same exascale high performance computing system. This workflow can be commonly applied to a broad variety of scenarios in which simulation campaigns play an important role. Unlike single simulations, campaigns run several simulations in parallel. This is interesting for finding an optimum operation point of a machine like a particle accelerator or a fusion reactor, quantifying the variation in model predictions e.g. in climate research, or discovering new capabilities e.g. in materials or drug design. Such campaigns produce huge amounts of high-fidelity data at high rates. File system bandwidth and capacity prohibit retaining all of that data, usually requiring orders of magnitude in reduction before it can be stored. We demonstrate a workflow that streams the data from multiple simulations to train a frontier model in parallel, thereby maximizing data throughput. We demonstrate that this can use a much larger amount of high-fidelity data as training data for the model compared to storage in files, essentially replacing file IO by in-transit training of large-scale frontier models. This will allow to create AI-informed feedback loops to steer simulation campaigns while reducing the overall computation time needed and provide high-fidelity frontier models for reuse by the scientific community.
\end{abstract}


\section{Introduction}
\sectionGuide{1 page?}{Sunita,Micha,Jeffrey,Richard}

\begin{figure}[bp!]
    \centering
    \includegraphics[width=0.49\textwidth]{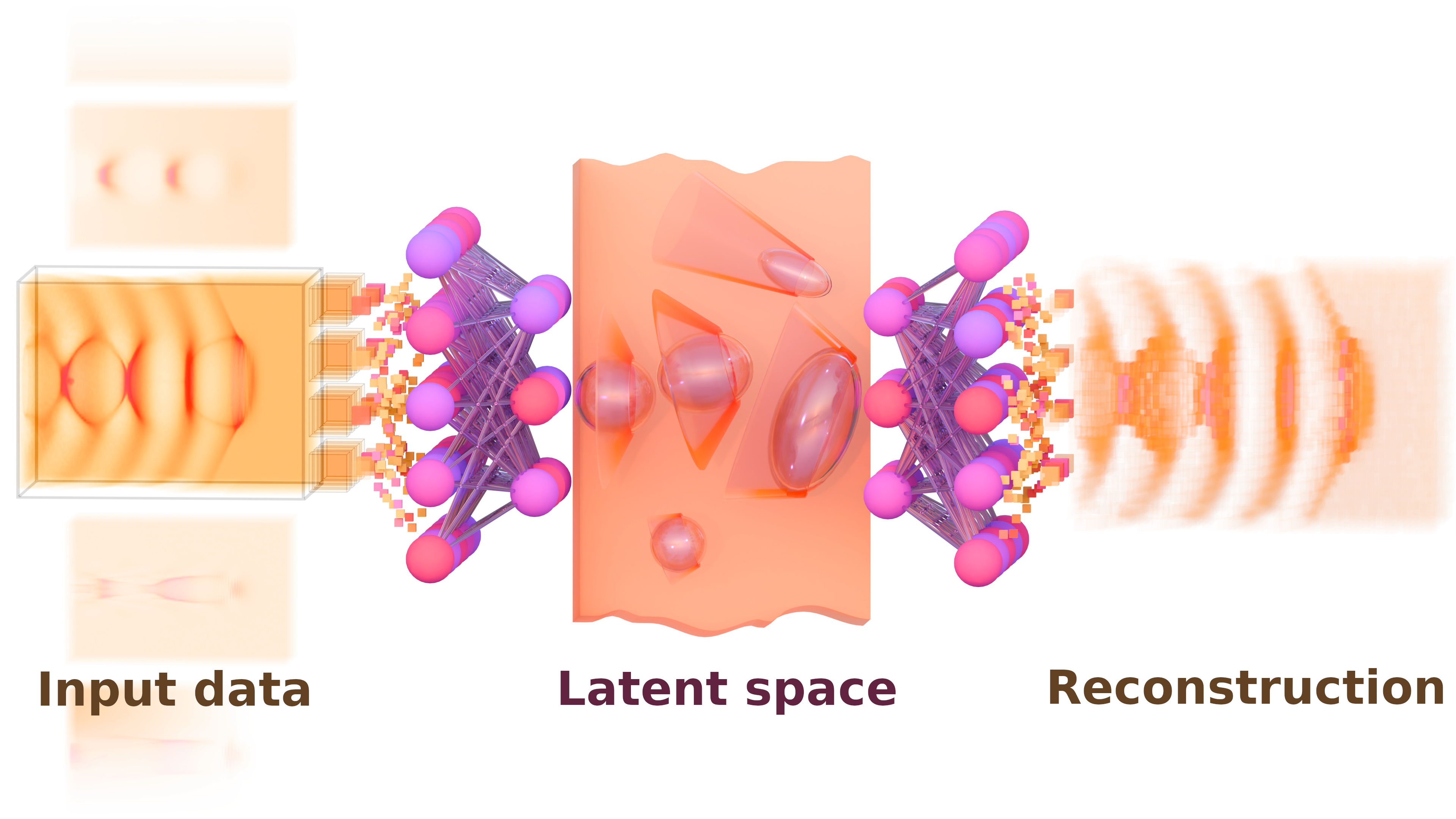}
    \caption{Impression of our parallel streaming and training workflow and the feature-extraction principle for the case of \gls{lwfa} simulations: data from multiple particle-in-cell simulations is streamed into a \gls{ml} training to learn a latent space representation of the plasma dynamics that later can be used to reconstruct the electromagnetic fields.}
    \label{fig:3DRecycler}
\end{figure}

Exascale simulations produce high-fidelity data at unprecedentedly high rates that exhibit dynamical features central to the understanding and control of complex systems. 
Applications include modeling weather, predicting materials properties or finding optimal operation points of fusion reactors. 
In this work, we take the example of \gls{lwfa} as an advanced accelerator technology for producing brilliant, high-current electron beams. 
We demonstrate how to retain information contained in high-fidelity data streaming from multiple simulations for in-transit machine learning to predict critical operation data based on the plasma dynamics.
% \atjk{Could be easily misunderstood! We do not `retain high fidelity data', nor do we 'retain' data streaming.}

Building on previous work on using a single simulation as a data source \cite{Kelling_IPDPS_2025}, in this work we extend the%our 
approach to using multiple simulations as data sources in parallel (depicted in Fig.~\ref{fig:3DRecycler}). 
% This approach enables us to cover a large parameter space in parallel as a critical technological step to implement feedback into covering that parameter space optimally. 
% Too convoluted, "covering parameter space" getting repeated.
% \vp{This approach is a critical technological step to implement feedback into covering a large parameter space in parallel optimally.} 
This parallel approach is a critical technological step that enables both 
large-scale parameter space coverage and adaptive feedback for optimized 
exploration.
When considering applications such as \gls{lwfa}, these spaces quickly become large and critical operation parameters usually exhibit a nonlinear dependence on simulation input parameters. 
It is thus vital to retain the data from multiple simulations in a single machine learning model that over time becomes a highly efficient representation of the system dynamics in the parameter range covered.

% This model can in turn be used to inform \vp{whom?} where operation critical information is missing and to decide which simulations to run next while interpolating data in well-covered areas of the parameter space. 
This model can, in turn, be used to identify where operationally critical information is missing and to determine which simulations to run next, while interpolating data in well-covered areas of the parameter space.
We call this approach ``The Recycler'', as it uses all relevant high-fidelity simulation data to create a surrogate model of all simulation data that can replace future simulation runs by querying the model while providing information on what simulation data is still missing.

Our approach demonstrates the convergence of large-scale parallel AI training and simulation campaigns aimed at discovery science and operation optimization on a single high performance computing system. 
It minimizes the loss of high-fidelity simulation data through in-transit
training of multiple simulations in parallel, \emph{enabling a dataset, albeit
ephemeral, large enough to train frontier models.}
This is a critical step towards
reducing the amount of simulations needed for covering large parameter spaces by
model inference - both by replacing simulation runs and optimizing parameter
space coverage.
%\atjk{Here MB goes already into feedback for steering future sims - can we keep this or is this too beyond the scope of this paper? }

The resulting model can be used for future inference by users without access to high performance compute systems, thus providing a single faithful, easily accessible, high-fidelity representation of many simulations. 
Our approach is universal in the sense that it is application and simulation agnostic. 
With \gls{lwfa} we have deliberately chosen a real-world application that demands extensive coverage of a complex, highly nonlinear parameter space, thus demonstrating the practical advantage of our workflow in comparison to previous work using single simulations as a data source.

The main results of our work are summarized below. We
{
\begin{itemize}
 \item report on the first in-transit machine learning on high-rate, high-fidelity data from multiple simulations running parallel to the neural network training on the same high performance computing system,
 \item demonstrate a complete in-situ workflow of running data reduction and neural network training parallel to batches of laser plasma simulations running in parallel that function as the sources of training data,
 \item discuss the challenges of scheduling multiple simulations and data-parallel machine learning on the same high performance computing system so that the compute is optimally used,
 \item confirm the advantage of in-transit learning of data from multiple simulations in parallel to retain high-fidelity data at high rates to circumvent the restrictions of limited throughput when using the file system for intermediate storage of training data and
 \item evaluate the reconstruction quality of application-critical data by model validation, thus exemplifying the benefit of using machine learning models as data representations over file output.
\end{itemize}
}

\section{Related Work}
\subsection{Laser Wakefield Acceleration} \label{sec:secLWFA}

\sectionGuide{2 paragraph = 0.5 column}{Richard}
{
}

\gls{lwfa} was first proposed by Tajima and Dawson in 1979~\cite{TajimaDawson1979}.
In an \gls{lwfa}, an intense laser pulse drives a plasma wake. 
This provides a highly compact accelerator design with acceleration gradients that are orders of magnitude higher than those of conventional accelerators.
A key challenge in \gls{lwfa} is controlling the injection of electrons into the wakefield.
One established approach is ionization injection, in which the plasma is doped with a small amount of a high-Z gas, such as nitrogen.
When the laser intensity becomes high enough, the inner-shell electrons of the dopant atoms are freed and can be trapped in the wakefield for further acceleration.
If the injection process stops due to evolving laser-plasma dynamics, it is referred to as \gls{stii}~\cite{HafzSTII2016}. 
Our setup utilizes this mechanism, which yields short, high-quality, high-current electron bunches~\cite{couperus_demonstration_2017}.

The simulation setup for this study is based on the experimental conditions of the DRACO laser experiments at HZDR\cite{SchrammDRACO2017}.
A \qty{30}{\fs} laser pulse with a \qty{20}{\um} spot size and a central wavelength of \qty{800}{\nm} is shot at a gas mixture consisting of \qty{96}{\percent} helium and \qty{4}{\percent} nitrogen.
The plasma density is based on experimental measurements and provides a constant electron density of approximately \qty{4e18}{\per\cubic\cm} over a distance of about \qty{1.4}{\mm}.
Since the laser pre-pulse is strong enough to ionize most of the plasma before the main pulse arrives, we can assume that the gas is pre-ionized.
Before the drive laser arrives, only the two innermost electrons of the nitrogen atoms remain bound. 
These electrons are used for self-truncated ionization injection~\cite{irman_improved_2018}.

\subsection{The Technical Challenge: Exascale Simulation Campaigns vs. Data
Storage and Bandwidth\label{subs:technical-challenge}}

\sectionGuide{1 column}{Jeffrey}
{
\begin{itemize}
 \item ~
\end{itemize}
}

When running campaigns of large-scale simulations, the final results are often derived from only the final outputs of simulations. 
Information about intermediate simulation states and thus about the process creating that final result is not stored at all or is stored only in a heavily reduced form. 
It would be advantageous to retain more information about the process that gives rise to the end result in order to better understand the physical process through expert analysis. 
This allows for the development of predictive models of the process, which could help approximate the results given the input parameters. 
In the end, this could help reduce the number of simulations required to efficiently cover a parameter space.

Effective data reduction with classical methods requires detailed knowledge about which parts of the simulation states primarily drive the physical process to cause the final outcome, i.e. it requires an understanding of the complex system which the data would help to obtain in the first place. Unsupervised or self-supervised \gls{ml} methods, such as \glspl{vae}~\cite{KingmaWellin2013_AutoEncodingVariatioBayes} or \glspl{mae}~\cite{HeChenXieLiGirshi2021_MaskedAutoencoScalableVisionLearners}, can help by capturing the structure of data and relations between features. In \glspl{jepa}~\cite{AssranDuvalMisraBojanoBallas2023_SelfSupervisLearningImagesJointEmbeddin}, this feature extraction can be steered towards capturing features which are most relevant in describing the physical system over features which are just perceptually prevalent in the dataset~\cite{MaesLidecScieurLeCunBalest2026_LeworldmStableEndEndJointEmbeddin}.

State-of-the-art self-supervised training workflows are built around training sets available in permanent storage. With 3D high-resolution simulation data, this becomes infeasible quickly, beyond a very small set of sparsely sampled runs. Training of frontier models requires a much larger number of training examples, by definition, to provide a good basis for generative performance and generalization across an entire research domain. Here, streaming simulation data directly into model training, bypassing the file system, offers a way out, but comes with a different set of challenges:
\begin{itemize}
    \item When training data is fed into a model sequentially, instead of in a randomized order, the problem of catastrophic forgetting~\cite{McClosCohen_1989CatastroInterferConnectiNetworksSequentiLearningProblem} arises: the model will be prone to over-fit on the currently streaming simulation or even on each successive transient simulation state. Strategies like \gls{er}~\cite{ChaudhRohrbaElhoseAjanthRanzat2019_TinyEpisodicMemoriesContinuaLearning} have been developed, which address the problem by keeping a sparse random selection of earlier training samples in buffers to mix into new batches. With high-dimensional data, the size of these \gls{er} buffers is limited.
    \item Simultaneous training on multiple running simulations, to increase data diversity in the global stream, creates workflow challenges: The consuming training code must find available simulation (producer) streams. 
    This is easier when a workflow is tightly coupled or static, or when the training data can be communicated via a central broker~\cite{BalinSiminiSimpsoShaoRigazzEllisBecker_2023SituFrameworCouplingSimulatiMachineLearningApplicatCfd}. 
    When simulations are driven not by model training, but by an external optimization loop, discovery of streams and simulation status must be dynamic.
    When simulations produce data in $\sim$\,\si{\tera\byte\per\s}, no broker can be employed for the data and all coordination between simulations and training must be through well-defined metadata.
    \item Ingesting multiple large simulations simultaneously also requires high data throughput on the training side, necessitating large-scale data-parallel training to keep up with data production.
    \item Training with very large batch sizes can hurt generalization performance of a model~\cite{KeskarMudigeNocedaSmelyaTang2017_LargeBatchTrainingDeepLearningGenerali}, which can be addressed by using small batches for warmup~\cite{GoyalDollarGirshiNoordhHe2018_AccurateLargeMinibatcSgdTrainingImagenet} and annealing learning by batch size, instead of learning rate~\cite{SmithKinderLe2018_DonTDecayLearningRateIncrease}.
\end{itemize}

\section{Methodology}
\subsection{Particle-in-cell simulation and its setup} \label{picsim}

\sectionGuide{2-3 paragraphs = 0.75 column}{Richard}
{
%\begin{itemize}
%    \item state that particle-in-cell (PIC) is the preferred method for modeling nonlinear and relativistic plasmas, such as in LWFA
%    \item briefly explain the particle-in-cell algorithm as cycle between: force-intrepolation onto particles, particle push based on forces, current deposition based on particle motion and field update based on current 
%    \item State that PIConGPU  is an open-source, 3D3V PIC code developed at HZDR used for all simulations. 
%    \item state that it is portable via alpaka (supports NVIDIA, Intel, AMD, OpenPOWER)
%    \item Point out that PIConGPU is the primary data source for the Recycler workflow by providing field/mesh data via openPMD streaming.
%    \item state exact technical simulation setup
%\end{itemize}
}

The \gls{pic} method~\cite{Hockney1988} is the most commonly used algorithm for modeling nonlinear and relativistic plasmas, such as those encountered in \gls{lwfa}.
\gls{pic} excels at capturing the intricate, multiscale dynamics of these systems through a cyclic process. 
First, electromagnetic fields are interpolated from the mesh to compute forces on particles. 
Then, particles are pushed based on these forces to update their positions and momenta. 
Next, the resulting currents are deposited back onto the mesh. 
Finally, the electric and magnetic fields are updated using Maxwell’s equations.

For all of the simulations, we use PIConGPU, an open-source, fully relativistic, 3D3V \gls{pic} code developed at HZDR  \cite{Bussmann2013}. 
PIConGPU is highly portable across diverse architectures via the alpaka library and supports NVIDIA, Intel, AMD, and OpenPOWER systems \cite{Zenker2016}. 
It serves as the primary data source for our Recycler workflow by providing field and mesh data via openPMD streaming \cite{Poeschel2022}.

\begin{figure}
 \centering
 \includegraphics[width=\linewidth]{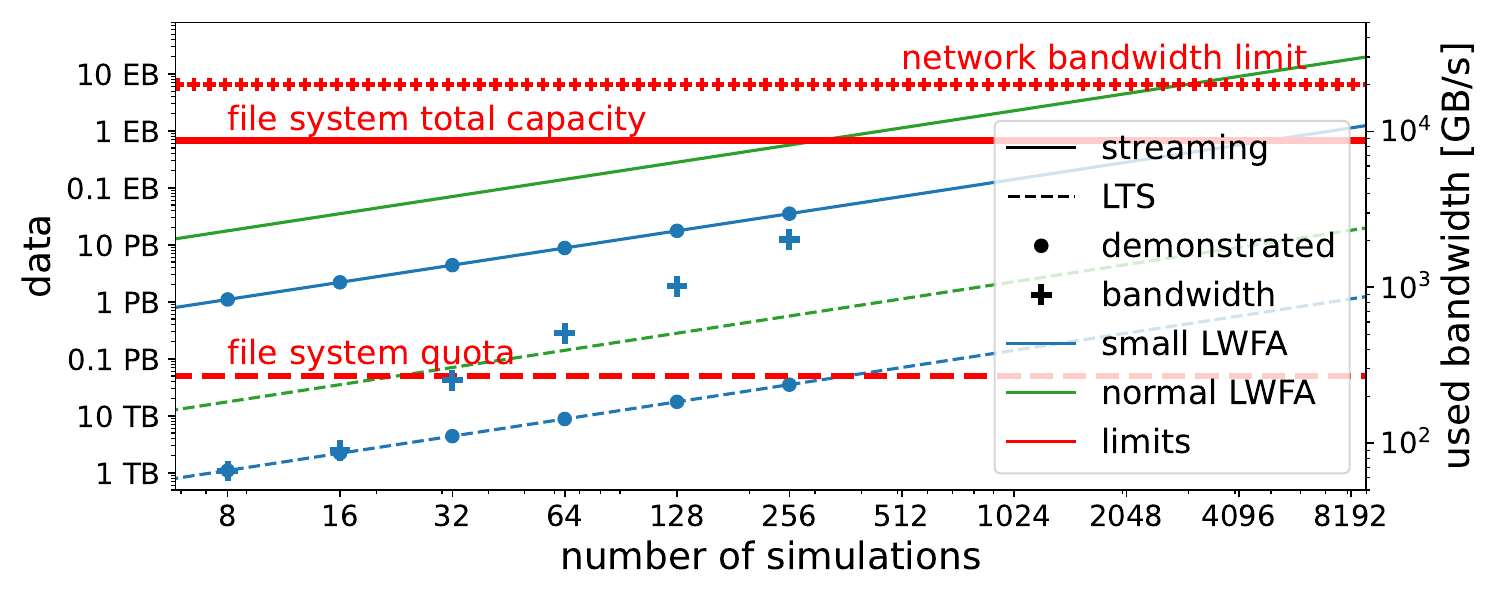}
 \caption{\label{fig:FSlimits}
 Scaling of data handled by our streaming-based Recycler method:
 Dots and plus signs represent the measurements from this publication. 
 Blue indicates streams and \gls{lts} from the small-scale \gls{lwfa} simulation we used here. 
 With our scaling run, we already reached the project's quota limit. 
 Note that streaming data is not written to disk and is just compared to the file system capacity as if it were written to disk. 
 In green, scaling for a better resolved \gls{lwfa} simulation is plotted. 
 By doubling resolution, its output increases $16\times$ and would thus already exceed quota limits for \gls{lts}.
 The used network bandwidth (plus signs) shows the average bandwidth used during the scaling runs. 
}
\end{figure}

Several PIConGPU simulations are executed in parallel on 16 MI250X GPUs each. 
Each simulation uses a $416 \times 2304 \times 416$ mesh that propagates alongside the driver laser through a few millimeters of plasma. 
This is a factor $2^4=16$ smaller than commonly used in our high-fidelity \gls{lwfa} simulations as this setup was considered a demonstrator and long-term storage (see Sec.~\ref{subsubsec:ER}) would have otherwise exceeded file system quotas (see Fig.\ref{fig:FSlimits}). Our \gls{ml} training workflow nevertheless transfers, as larger inputs could be ingested into training, putting more pressure on network bandwidth and pre-processing compute, as well as limiting the number of \gls{lts} samples further. However, higher fidelity simulations could also drive the exact same training: Higher resolution primarily yields more accurate simulations. Global outcomes do not necessarily have to be represented at the same resolution and hence a model may learn a practically useful approximation by ingesting accurate simulation states at a lower resolution. 

The \textit{mesh} data, which includes electric, magnetic, and current vectors, is streamed via openPMD-api \cite{openPMDapi} and ADIOS2 \cite{adios2} for real-time processing in the \textit{Recycler} workflow.
%\atrp{shall we italicize the word 'mesh' to emphasize it (in the interest of people outside of computational physics field, used often later and throughout)? or call here `field data' which is what it is and will be understood by anyone? Likewise, `iteration' in ML context is something entirely different (I suggest we use `simulation time-step' to make our reference clearer to everybody, a specialization-agnostic common vocabulary ;)).}

\begin{figure*}[t]
    \centering
    \includegraphics[width=0.92\textwidth]{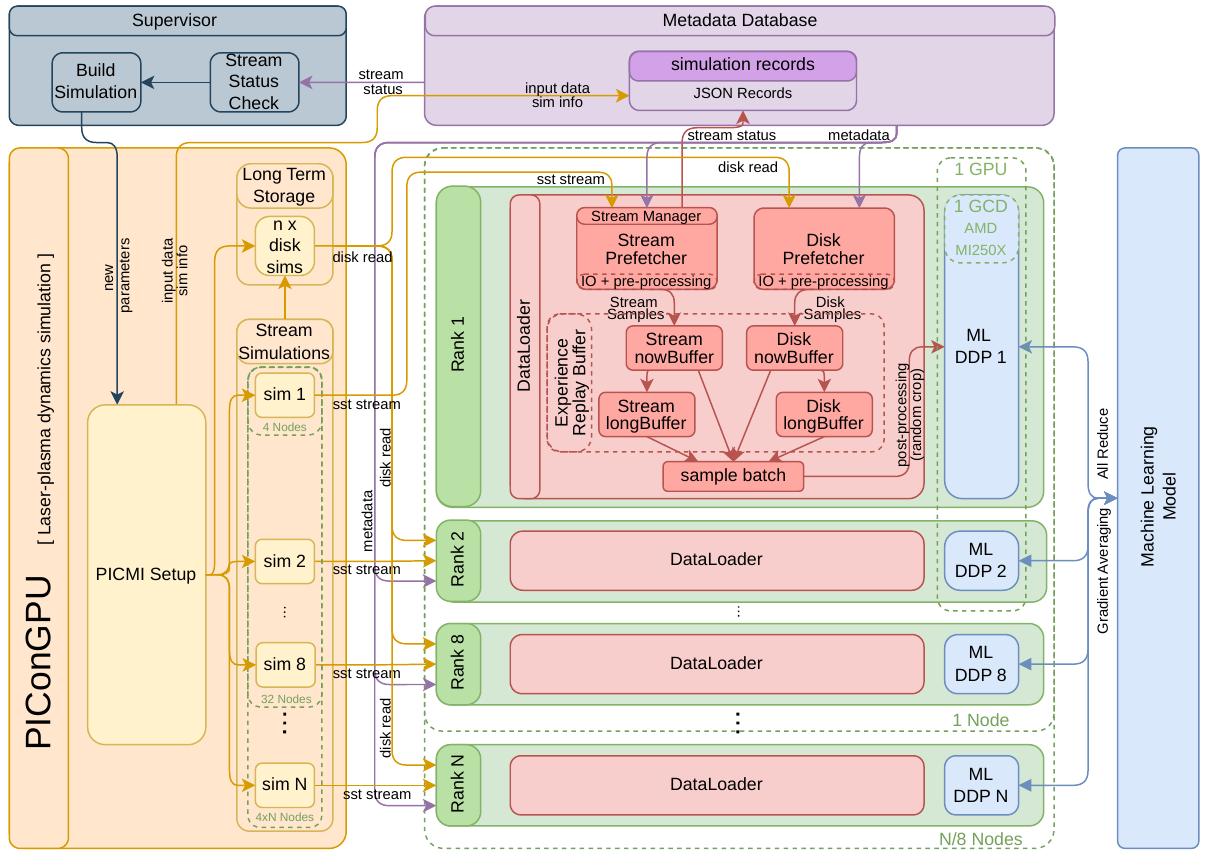} 
    \caption{The Recycler workflow. PIConGPU simulations (left) stream their fields over \gls{sst} to the ranks of the data-parallel training (center), each of which has its own DataLoader with prefetchers and experience replay buffers for the stream and for long-term storage on disk. The batches it produces train one replica of the model per rank, and the gradients of all ranks are averaged in an all-reduce (right). The metadata database (top) coordinates simulations, ranks and the supervisor, which replaces a simulation when its stream has ended.}
    \label{fig:The-Recycler}
\end{figure*}

\subsection{The Recycler} \label{openPMD_streaming}
%\subsubsection{Overview}
Centrally, the Recycler separates data loading from the training loop (Fig. \ref{fig:The-Recycler}). Data loading, which includes sample-selection based on metadata, disk and stream IO, pre-processing and buffering, is the task of the DataLoader, which exposes a synchronous iterator interface to the training logic. 
The iterator generates batches on demand, mixing samples from two data channels, disk and stream, and performs
post-processing.
Each data channel has asynchronous prefetchers and a separate \gls{er} buffer \cite{Kelling_IPDPS_2025}.
Prefetching and sample picking are guided by metadata. 
The training logic consumes batches from the iterator. In the presented setup, it prefetches batches asynchronously.

\subsubsection{Prefetching}
Each data channel is served by an asynchronous worker, called a prefetcher, which fills a prefetch queue. There is one prefetcher for the disk channel and one for each stream.
IO from both disk and stream is handled using openPMD \cite{Poeschel2022}, which also provides run-level metadata for checks, e.g., of mesh shapes and available simulation time steps, as input to dynamic transformations. A prefetcher performs IO and pre-processing asynchronously in separate threads, so that IO and compute overlap.
To actually enable asynchronous data loading in the background without stalling any other workers, the openPMD-api had to be extended with a more fine-grained awareness of the Python \gls{gil}.
The pre-processing pipeline is configurable and handles data cropping, downsampling, normalization and data format conversions.

From disk, samples are picked randomly across simulations and time steps, i.e., a random simulation and then one of its time steps, based on the availability information provided by the metadata database.

From stream, samples are received using the \gls{sst} engine of ADIOS2 \cite{Eisenhauer2024}. The stream channel is run by the stream manager. It periodically queries the metadata database for streams and attaches to those it is responsible for, creating a prefetcher for each of them. If an attachment fails, e.g., because the simulation has not started to publish yet, it is repeated in later queries up to a configurable number of attempts. When a batch is generated, the stream manager collects the available samples across the prefetch queues of its streams without having to wait for any of them. A stream whose simulation has ended is retired once its prefetch queue is empty, and it is not attached again.

\subsubsection{Metadata Database}
Simulations are described in a metadata database — a NoSQL document database on disk — based on the PICMI metadata~\cite{picmi_standard} of the simulations. A database abstraction layer provides a filter that the DataLoader uses to select which simulations to read, e.g., by type of data source, mesh shape.

The stream manager also communicates through this database. A stream manager records the state of each of its streams in the record of the simulation: \textit{waiting} when it attaches, \textit{running} when it gets its first sample, and \textit{ended} when the stream is retired, or \textit{unavailable} if no sample was ever received after a set number of tries. Three parts of the workflow read these states. The disk channel checks if a stream has reached its final state and only then offers its outputs. The stream assignment treats the rank of a finished stream as free again. The supervisor builds and launches a new simulation when a stream has ended. 

In data-parallel training, every rank has its own DataLoader and stream manager, and the database also serves to distribute the streams among the ranks. Every stream manager goes through the streams in the order of their registration and assigns each stream to the rank that has the fewest streams at that point. A rank counts as free again shortly after its previous stream has reached its final state, so that the stream of a replacement simulation goes to the rank whose stream has just ended. Every rank computes this assignment independently from the same records, using the same deterministic rule (no randomness is involved). As a result, all ranks agree on the owner of every stream without needing to communicate with each other. The assignment of a stream never changes.

\subsubsection{Batch Sampling}
To generate a batch, samples from the prefetch queues are staged for insertion into the \gls{er}. Each data channel has its own \gls{er} buffer (Sec. \ref{subsubsec:ER}). Each channel contributes a configurable share of the batch, which is filled from the staged samples first, then by random selection from the \gls{er} buffers.
Samples may be picked once, with the iterator blocking until enough samples are available, or picked multiple times to fill the batch.
We do the latter to avoid a slow buffer warmup, which works because picked samples are augmented by the post-processing. 
While its buffer fills, the disk channel picks samples repeatedly, and it also completes the batch when the stream channel cannot yet supply its share. The stream channel itself never repeats a sample within a batch. The samples picked for the batch are post-processed by a configurable pipeline, which in this workflow is a random crop. Finally, the staged samples are inserted into the \gls{er} buffer.

\subsubsection{Experience Replay}
\label{subsubsec:ER}
\gls{er} is a mechanism to avoid catastrophic forgetting in model training when the data arrives sequentially instead of randomly.
It keeps random older samples from the stream to mix in with recent samples (Sec. \ref{subs:technical-challenge}). As in \cite{Kelling_IPDPS_2025}, we completely decouple \gls{er} from training and move it into the DataLoader as a buffering mechanism. The \gls{er} buffer consists of two components: the now-buffer, which behaves as a ring-buffer, and the long-buffer, which contains random older samples. It is updated as follows (see also, \cite{Kelling_IPDPS_2025}):
\begin{itemize}
    \item New samples, staged from the prefetch queues, are pushed to the front of the now-buffer.
    \item When the now-buffer overflows, the samples that have been picked most often leave it. A fraction of them (0.3 by default, rounded up to a whole number of samples) is promoted at random to the long-buffer, and the others are discarded.
    \item When the long-buffer exceeds its size, random samples are removed from it.
\end{itemize}

Because samples arrive one at a time, each sample that leaves the now-buffer is promoted, so the long-buffer fills quickly and then holds a random selection of older samples. 
Sampling from \gls{er} is random, with configurable fractions of the samples taken from the now-buffer and the long-buffer. By default the fractions are proportional to the sizes of the two buffers.

By using a disk channel in addition to the stream \gls{er}, the Recycler adds \gls{lts} as a third replay layer. At the end of a run of the workflow, the database records of its simulations can be passed on to the next run to serve as \gls{lts}. The disk channel also uses \gls{er}, but only a now-buffer, for caching in conjunction with data augmentation. A long-buffer is not needed because data is loaded from disk at random.

\subsubsection{Overall Workflow}
In this workflow (Fig. \ref{fig:The-Recycler}), the simulations are run under the control of either an acquisition function for parameter space exploration or optimization, or as part of an active-learning loop controlled by the model being trained. Simulations are registered with the metadata database when they are launched. 
A simulation first provides its stream, and once the stream is marked as ended, the outputs that it has written on disk become available as \gls{lts}. The Recycler runs alongside, connects to the active streams and loads \gls{lts} samples. The model consumes the batches yielded by the DataLoader.

\sectionGuide{2 pages}{Jeffrey, Franz, Nikhil}
{
\begin{itemize}
 \item technical description of architecture
 \item overview
  \begin{itemize}
      \item centrally, The Recycler separates data loading, including sample-selection based on meta data, disk and stream IO, as well as pre-processing and buffering, from the training loop using the DataLoader which exposes a synchronous iterator interface to the training logic
     \item the iterator generates batches on demand, mixing samples from disk and stream (data channels) and performs post-processing
     \item each data channel has an asynchronous prefetcher and a separate experience replay buffer (ER)~\cite{Kelling_IPDPS_2025}
     \item prefetchers/sample-picking are guided by meta-data 
     \item the training logic consumes samples from the iterator. in the presented setup, it prefetches batches asynchronously.
  \end{itemize}
 \item prefetching (par or subsubsection)
  \begin{itemize}
   \item for each data channel an asynchronous worker thread (prefetcher) fills a 
   prefetch queue.
   \item IO from both disk and stream are handled using openPMD (providing run-level meta data for check or as input to dynamic transformations)
   \item \emph{from disk:} samples are picked randomly across simulations and time steps, based on availablity information provided by the meta datadb
   \item \emph{from stream:} using adios SST for streaming \atfranz{maybe we could add some detail here}
   \item include asynchronous IO and pre-processing (configurable pipeline: cropping, down-sampling, norm-scale) in the prefetcher for IO/compute overlap). \emph{here, point out, that during our work we discovered, that openpmd-api did not realse GIL during IO and fixed the issue upstream to enable IO/compute overlap} \atfranz{maybe we can add some detail or reasoning here and cite the PR, Answer Franz: Check my comment on the same subject some pages earlier. Details about fixed bugs are too low-level for a research paper, but we can frame GIL-awareness as a new feature needed in an IO library to be used for asynchronous background data loading.}\jk{done}.
   \item \atnikhil{add details about the stream manager}
  \end{itemize}
 \item meta data db
 \begin{itemize}
     \item database abstraction payer for filtering
     \item ... no details on db, just: (noSQL-docmument DB) on disk based on picmi simulation meta data. 
     \item \atnikhil{stream manager communication goes here}
 \end{itemize}
 \item batch sampling
  \begin{itemize}
   \item samples from the prefetch queues staged for insertion into the ER
   \item each data channel has its own ER (ref later explanation)
   \item the batch is filled from the staged samples first, then by random selection form ER buffers.
    \begin{itemize}
     \item samples may be picked once, with the iterator blocking until enough samples are available, or picked multiple times to fill the batch \emph{this is what we do avoid slow buffer warm-up and works because picked samples are augmented by the post-processing}
    \end{itemize}
   \item samples picked for the batch are post-processed by a configurable pipeline (random crop, (stacking))
   \item staged samples are inserted into ER (-> next subsubsection)
  \end{itemize}
 \item experience replay
  \begin{itemize}
      \item a mechanism to avoid catastrophic forgetting in model traiing when training data arrives sequentially instead of randomly
      \item ER keeps random older samples from the stream to mix in with recent samples~\cite{Chaudhry2019OnTE}
      \item we completely decouple ER from training~\cite{KellinBoleaBussmaCheckeDebusEbertEisenh_2025ArtificiScientisTransitMachineLearningPlasmaSimulati}, moving into the dataloader as a buffering mechanism
      \item The ER buffer consists of two components: now-buffer, which behaves as a ring-buffer, and long-buffer, which contains random older samples.
      \item brief description of update mechanism: (also see~\cite{KellinBoleaBussmaCheckeDebusEbertEisenh_2025ArtificiScientisTransitMachineLearningPlasmaSimulati})
      \begin{enumerate}
          \item new samples (staged from prefetch queue) are are pushed to the front of the now-buffer
          \item when the now-buffer overflows, overflow samples move into the long-buffer, as long as space is available to fill bufffers quickly
          \item when the long-buffer is full, only a fraction (\texttt{nowLongPromotionFraction: float = .3}) of overflowing samples are promoted at random. each promoted sample replaces a random sample in the long buffer.
      \end{enumerate}
       \item sampling from ER is random, with configurable fractions of samples taken from now- and long-buffer (\texttt{nowSamplingFraction: float = None}, i.e. the ratio is ratio of now-buffer size/long-buffer size).
       \item the recycler using a disk channel in addition to stream ER, adds long-term-storage (LTS) as a third replay layer. \atnikhil{mention the addition of previous runs' output to the DB for use in the next run in this context.}
       \item as a minor detail we could add, that the disk-channel also uses ER, but only a now-buffer (long-buffer size =0), for caching in conjunction with data augmentation. long-buffer is not needed, because data is loaded from disk at random.
  \end{itemize}
  \item overall workflow (in theory) + summary
   \begin{itemize}
    \item simulations are run, controlled by either an acquisition function for parameter space exploration or optimization, or as part of an active-learning loop controlled by the model being trained (in our case we have an "acquisition function for parameter space exploration", which is just random: do not mention this here, only in the Frontier setup)
    \item simulations are registered with the meta-data DB (first streams, then final outputs marked done -> LTS)
    \item recycler runs, connects to active streams, loads LTS samples
    \item model consumes batches yielded by DataLoader.
   \end{itemize}
\end{itemize}
}

\subsection{Machine learning} \label{machine_learning}

\begin{figure*}[t]
    \centering
    \includegraphics[width=\textwidth]{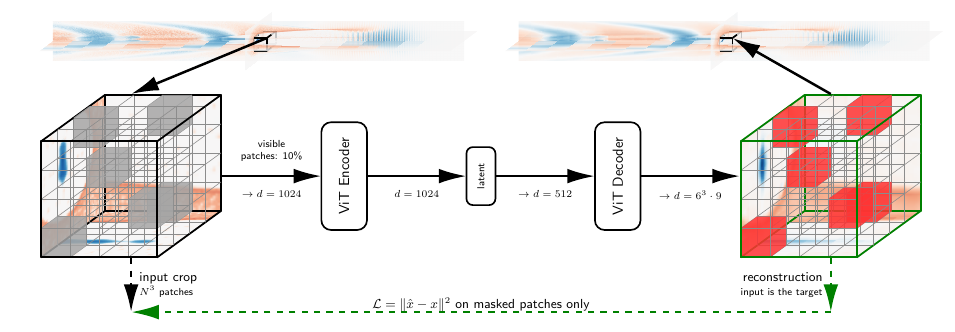} \\
    \caption{Summary of data flow during training and self-supervised objective: The loaded/streamed and buffered samples contain a region of interest from the simulation of $144 \times 1536 \times 144$ cells. As training samples, random regions of $48^3$ cells are cropped, which are split into $8^3$ patches, which get projected into the 1024-dimensional embedding space of the \gls{vit} encoder. Each latent token is projected down into the 512-dimensional embedding of the \gls{vit} decoder. For \gls{mae} training, only a random \qty{10}{\percent} of input patches (gray cubes, $N=51$) are passed to the encoder. The decoder receives the encoded tokens and reconstructs all tokens, with the reconstruction loss being applied only on the complementary \qty{90}{\percent} masked patches.}
    \label{fig:mae}
\end{figure*}

To process the high-fidelity streaming data without the need for manual labeling, we use \gls{ssl}. The primary goal of this approach is to extract the underlying patterns and nonlinear relations that describe the plasma dynamics efficiently, mapping the high-dimensional simulation space into a compact latent representation.

From a data-management perspective, this learned representation serves as a dense feature extraction mechanism. Rather than focusing on traditional storage compression, the \gls{ssl} objective is designed to distill the raw physical fields into a latent space that preserves details critical to the physics while discarding redundant or noisy information. This latent space provides a structured foundation for downstream models, specifically latent surrogate models that can predict simulation outcomes without requiring the full execution of the \gls{pic} code.

We employ a Vision Transformer architecture adapted to process 3D spatial volumes, expanding the input space from standard image channels to nine physically coupled channels corresponding to the three components of the electric field, magnetic field, and current density: $\vec{E}=\{E_x,E_y,E_z\}$, $\vec{B}=\{B_x,B_y,B_z\}$ and $\vec{J}=\{J_x,J_y,J_z\}$. By training the model via masked autoencoding followed by cross-attention, we aim to steer the network beyond learning simple pointwise correlations and consequent interpolations. The intent of this objective is to encourage the model to capture the spatial dependencies and phase-locking characteristics of laser-plasma interactions, which are fundamentally governed by Maxwell's equations \cite{maxwell1873treatise}.
We implement our model based on the MuViT framework~\cite{MantesMannoWeiger_2026MuvitMultiResolutiVisionTransforLearningAcrossScales,muvit_github}. Built upon the \textit{x-transformers} library\cite{xtransformers_github,vaswani2017attention,dosovitskiy2021an}, the model utilizes a \gls{mae} objective~\cite{HeChenXieLiGirshi2021_MaskedAutoencoScalableVisionLearners} as the self-supervised target. In this setup, a significant portion (\qty{10}{\percent} in our case) of the input data patches are masked, and the model is tasked with reconstructing the original field values of these missing regions (see Fig.~\ref{fig:mae}).

Crucially, this \gls{mae} objective is chosen because it forces the encoder to develop a global understanding of the plasma wake and laser-plasma interaction, rather than relying on simple local interpolation. To reconstruct a masked region—such as the accelerating cavity—the encoder must infer the state of the system from the visible patches, effectively extracting high-level physical features rather than relying on simple local interpolation. As discussed in Sec.~\ref{sec:MLresults}, this results in a latent space where physically relevant information is encoded in a way that remains linearly accessible, allowing a simple linear probe to map these latent tokens to critical observables like the electron energy spectrum.

To integrate these capabilities into the in-transit workflow, we developed a specialized software module that wraps the MuViT implementation. This wrapper facilitates a seamless connection between the Recycler's streaming DataLoader and the training loop, ensuring that data-parallel GPUs are fed efficiently. Furthermore, this module provides a unified interface for hyper-parameter optimization and scientific analysis, including the latent \gls{pca} and volumetric visualizations used to evaluate the model's physical fidelity.

The encoder architecture is based on a \gls{vit} \cite{MantesMannoWeiger_2026MuvitMultiResolutiVisionTransforLearningAcrossScales}. From the input 3D volumetric data, crops of $48^3$ cells are partitioned into $6^3$-sized patches, yielding $8^3=512$ tokens, each of which is then projected into a latent embedding space. The encoder consists of 2 transformer layers with 8 attention heads and a hidden dimension of 1024. To reconstruct the masked voxels, the architecture employs a shallow decoder with a hidden dimension of 512. During training, a \qty{10}{\percent} masking ratio is applied to the input patches, forcing the encoder to capture global physical dependencies to successfully recover the missing data.
To account for the spatial nature of the simulation data, we employ 3D \gls{rope}~\cite{su2024roformer}. Because the DataLoader applies random spatial crops to the streaming meshes, the model must remain invariant to the absolute origin of a crop while retaining precise relative spatial awareness. \gls{rope} achieves this by applying a rotation to the query and key vectors in the attention mechanism, ensuring that the attention weights depend only on the relative distance between tokens.

To stabilize the model before exposing it to the high-rate simulation stream, we first perform a warmup pre-training phase using a small batch size on a representative, small on-disk dataset~\cite{keskar2017on}. This initialization phase prevents the instabilities often associated with early training on large batches and ensures that the model captures a basic representation of the plasma dynamics before shifting to in-transit learning. Once the model is stabilized, we transition to the streaming workflow and utilize a training-scaling approach characterized by annealing via batch-size increase~\cite{SmithKinderLe2018_DonTDecayLearningRateIncrease}. By gradually increasing the batch size throughout the training process, we maximize GPU throughput and stabilize the convergence of the frontier model, ensuring that the training keeps pace with the high-rate data stream provided by the simulations.

\sectionGuide{2 pages}{Jeffrey, Vedhas}
 {
 \begin{itemize}
  \item summary of self-supervised learning: extracting patterns and non-linear relations which describe input data efficiently
 \item perspectives: data compression (lossy but high compression, can be tuned towards important details of the physics), down-stream models, esp. latent surrogate models
  \item explain MAE as self-supervised objective (see introto linear probe in sec. \ref{sec:MLresults})
  \item describe ViT architecture. provide numbers (layer, dims)
  \item 3d ROPE~\cite{MantesMannoWeiger_2026MuvitMultiResolutiVisionTransforLearningAcrossScales}
 \item training-scaling approach with annealing by batch-size increase~\cite{SmithKinderLe2018_DonTDecayLearningRateIncrease}
   \begin{itemize}
       \item also include the warmup-pretraining with small batch size on a small on-disc dataset. \cite{KeskarMudigeNocedaSmelyaTang2017_LargeBatchTrainingDeepLearningGenerali}
   \end{itemize}
 \end{itemize}
 }

\subsection{Setup on ORNL Frontier system}  \label{frontiersetup}
%\atjk{Need review of this section}
A Frontier node has a 64-core AMD EPYC processor, of which 56 cores are available to jobs, \qty{512}{\gibi\byte} of memory, of which \qty{500}{\gibi\byte} (\qty{537}{\giga\byte}) is allocatable, 4 AMD MI250X GPUs that expose 8 \glspl{gcd}, and 4 HPE Slingshot network interfaces. 
Each simulation of Sec.~\ref{picsim} uses 4 nodes with 32 MPI ranks, one per \gls{gcd}. 
The model is trained with PyTorch \gls{ddp} training with eight ranks per node, one per \gls{gcd} with each rank given 7 CPU cores and tied to a single simulation stream.
\textit{N} simulations feed \textit{N} ranks on $N/8 + 4N$ nodes (\textit{N/8} nodes for training and 4 nodes per simulation for running the simulations), from 33 nodes for \textit{N=8} to \num{1056} nodes for \textit{N = 256}.

Our original plan was to run the simulations and the trainer as separate Slurm jobs, but on Frontier, processes of different jobs cannot establish \gls{rdma} connections over Slingshot %with each other and 
so \gls{sst} would fall back to slower TCP sockets. %The reason for this limitation is both a security constraint as well as a technical limitation.
For our current implementation we run all the simulations and the training as steps of a single job. Within the job, one job-wide virtual network identifier (\verb|--network=job_vni|) lets the steps reach each other via the \gls{rdma} data plane. {This setup circumvents not only the networking limitation but also a limitation of the scheduler: the simulations and the training have to run simultaneously, which separate Slurm jobs cannot guarantee. Within a single job, training starts together with the simulations, and replacement simulations start on freed nodes without queueing.} 

PIConGPU (0.9.0-dev) is built with HIP on ROCm 6.4.2 and Cray MPICH 9.0.1. The trainer uses PyTorch 2.13 bundled with ROCm 7.2 and runs in single precision. Both sides use openPMD-api 0.17.1 with ADIOS2 2.10. 

Each rank %of the training 
handles two kinds of traffic over Slingshot: the data stream from its simulation and the gradient exchange via DDP, both reaching the network through libfabric and CXI 
%in which DDP averages the gradients of all ranks at every step.
%Both reach the network through libfabric and CXI, its provider for Slingshot. 
{%The data stream uses the RDMA data plane of SST and the gradient exchange uses RCCL, AMD’s port of the NCCL, with aws-ofi-nccl plugin
The data stream uses SST's RDMA data plane and the gradient exchange uses RCCL with the aws-ofi-nccl plugin (v1.19.2, on Frontier~\cite{pytorch_frontier_docs}), a libfabric backend as named in \cite{Kelling_IPDPS_2025} to avoid the socket limit of the default backend. With both active in one process, the cross-node training hung at the first gradient exchange after a stream had been attached; disabling Slingshot eager messages resolved this\footnote{Documented workaround, \url{https://docs.cscs.ch/software/communication/nccl/\#known-issues} setting \texttt{FI\_CXI\_RDZV\_GET\_MIN}, \texttt{FI\_CXI\_RDZV\_THRESHOLD}, and \texttt{FI\_CXI\_RDZV\_EAGER\_SIZE} to 0 }. In a test with 16 ranks across 2 nodes, the configuration trained 1.8 times faster than the TCP socket backend.}

Each simulation publishes the electric field vector $\vec{E}$, the magnetic field vector $\vec{B}$, and the current density vector $\vec{J}$ for the entire volume every 10th of its \num{110000} time steps and writes the three fields at 11 randomly chosen iterations to disk. 
These 11 snapshots total \qty{138}{\giga\byte} and become \gls{lts}; streaming all steps would require \qty{138}{\tera\byte} per simulation.
%If the simulation were to write all the steps that were streamed it would take 138TB per simulation.
%\atjk{There are 2 abouts and 1 estimate with three numbers here. Is this really so uncertain, or are the numbers already correct within the prcision of their stated digits?}

The prefetcher reads a window of $144\times1536\times144$ cells, costing about \qty{1.15}{\giga\byte} per sample instead of \qty{12.6}{\giga\byte} for the full fields, then checks for non-finite values and normalizes via symmetric logarithm. Requesting only this window via openPMD, instead of reading the full fields and cropping afterwards, made the trainer run 1.6 times faster.

A batch has 32 samples per rank, 24 from stream and 8 from disk, with a random $48^3$ crop taken during post-processing. The stream \gls{er} buffer holds 4 samples in its now-buffer and 24 in its long-buffer. The disk now-buffer holds 9 samples. 
%Large buffers benefit the training but memory limits them.  
Together these buffers hold 37 samples and take \qty{42}{\giga\byte} per rank. With batch size of 32, a rank uses about \qty{59}{\giga\byte} at its peak, so eight ranks per node use about \qty{470}{\giga\byte} of the \qty{537}{\giga\byte} that a job can allocate, whereas batch 48 would need an estimated \qty{620}{\giga\byte}.

%DDP synchronizes all ranks at every step, because of which a stall in the DataLoader of one rank delays all of them. Two changes removed most of the stalls, the update from openPMD-api 0.17.0 to 0.17.1 with the GIL fix mentioned in Sec. \ref{openPMD_streaming}, and advising the kernel after each read from disk to drop the cached pages of the file. 
%In single-node tests using 8 disk-only ranks, the median iteration duration was 1.65s when neither optimization was applied, 1.14s with openPMD-api 0.17.1 enabled on its own, 1.05s when only the cache drop was used, and 0.26s when both changes were applied together. In addition to this, the disk prefetcher of a rank was throttled to fetch at most a sample every 40s which got the iteration time down to 0.22s.
\gls{ddp} synchronizes all ranks at every step, so a DataLoader stall on one rank delays all others. Two changes removed most stalls: updating openPMD-api to 0.17.1 with the \gls{gil} fix (Sec. \ref{openPMD_streaming}), and advising the kernel to drop cached pages after each disk read. In single-node tests with 8 disk-only ranks, median iteration duration was \qty{1.65}{\s} with neither optimization, \qty{1.14}{\s} with the openPMD fix alone, \qty{1.05}{\s} with cache-drop alone, and \qty{0.26}{\s} with both combined. Throttling the disk prefetcher to one sample every \qty{40}{\s} further reduced iteration time to \qty{0.22}{\s}.

When a stream ends, a supervisor (Fig. \ref{fig:The-Recycler}) builds and launches a simulation with new parameters to provide fresh samples to the DataLoader.
The randomized parameters for PICMI include the laser spot size $w_0$ (between \qtyrange{18}{22}{\um}), the normalized laser field strength $a_0$ (between \numrange{2.0}{3.0}), and the laser focus position, relative to the start of the simulation, $z_f$ (between \qtyrange{1.0}{2.5}{\mm}). 
The builds are done on the freed nodes and take a median of \qty{131}{\s}. 

The job is limited to \qty{5}{\hour} to cover the entire run time of a \gls{pic} simulation which took from \qty{2}{\hour} to \qty{4}{\hour} for \gls{lts} only runs. The jobs of the campaign, from 1 node trainer (8 ranks) to 32 node trainer (256 ranks) are called stages, each of which resumes from the latest model checkpoint and uses the completed simulations of all earlier stages as \gls{lts}.

\sectionGuide{2 par}{Nikhil}
{
\begin{itemize}
 \item give parameters used, like ER buffer sizes, batch size and briefly mentions reason for choosing them: usable memory capacity (more buffer=better for ML), max throughput on GPU
 \item also explain the issue with slingshot communication between nodes of different jobs being blocked, forcing us to have all sims+training in a single job, while the original idea was to have separate slurm jobs.
 \item do not take the length limit above too seriously
\end{itemize}
}

\section{Results}

\subsection{Weak-Scaling}
\begin{figure}
 \centering
 \includegraphics[width=\linewidth]{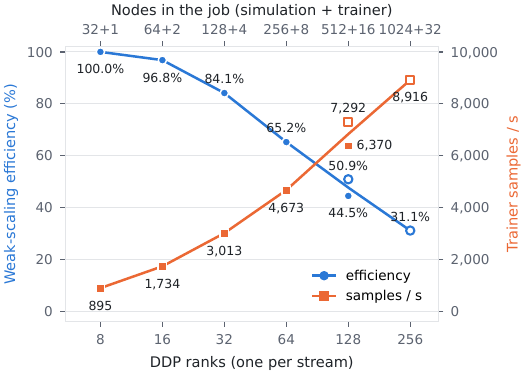}
 \caption{\label{fig:weak_scaling}
 Weak scaling of The Recycler on Frontier from 8 to 256 concurrent simulations with one \gls{ddp} training rank per simulation. The top axis gives the nodes of the simulations plus those of the training. Efficiency (left axis) is the mean iteration time with 8 \gls{ddp} training ranks divided by the mean iteration time with $N$ ranks. Throughput (right axis) is the number of training samples per second, i.e. global batch size divided by mean optimizer iteration time, where the global batch size is $N \times b$, with local batch size $b=32$.}
\end{figure}

We scale the number of simulations and number of \gls{ddp} training ranks together from 8 to 256. Each simulation feeds one \gls{ddp} rank, which trains on local batches of 32 crops. Ideal weak scaling would have a constant time per training iteration. Fig. \ref{fig:weak_scaling} shows two metrics, the efficiency relative to 8 \gls{ddp} ranks and the throughput in training samples per second, both defined in the caption. Runs with 8 to 128 simulations are chained stages as described in Sec.~\ref{frontiersetup}, while the second run with 128 simulations and the run with 256 simulations started from the initial checkpoint.

We see a decline in efficiency with growing number of ranks, owing to the cost of a global all-reduce operation to average gradients globally as part of \gls{ddp} training. Here, PyTorch is overlapping back-propagation compute and global communication to some extent, but is ultimately unable to hide the increasing communication cost. The efficiency stays close to ideal up to 16 simulations and falls to about \qty{31}{\percent} with 256 simulations. Overall, the total throughput increases steadily, albeit at growing cost: essentially 32 times the resources train about 10 times faster. Ultimately this demonstrates that our workflow can handle \emph{recycling} transient data from large, parallel simulation campaigns.

{The two runs with 128 simulations differ by \qty{12}{\percent} in their mean iteration time, a variance between runs for which we do not have an explanation yet. The other points, each of which is a single run, should be read with a similar uncertainty, which probably grows with the number of nodes. In the larger runs iteration time also increases over the course of a run. As all ranks synchronize in the gradient exchange at every iteration, this is unlikely to come from ranks drifting apart, and it points to something that accumulates over the run.}

{A single job here has to schedule and coordinate up to 256 simulations, their data streams, the metadata database, the supervisor and the distributed training on up to 1056 nodes, which all share the network and file system. Optimizing a workflow with this many interacting components is a new problem and their interplay is largely unexplored. Beyond tuning this workflow, other means to improve efficiency are to add other dimensions of parallelism to the training. With data parallelism alone, all N ranks hold a full replica of the model and take part in each gradient exchange. Pipeline and tensor parallelism instead shard one model replica over several \glspl{gcd}. We leave the combination of data, pipeline and tensor parallelism with \gls{ddp} to future work.}

\sectionGuide{1 col, 1 fig}{Nikhil}
{
\begin{itemize}
 \item ~
\end{itemize}
}

\subsection{Model Evaluation and Application: \gls{lwfa}} \label{sec:MLresults}

To verify that the self-supervised representation learned by the Recycler is physically meaningful, we evaluate the model through both reconstruction quality and its utility for downstream predictive tasks.

\subsubsection{Learned Representations}

\begin{figure}
 \centering
  \includegraphics[width=\linewidth]{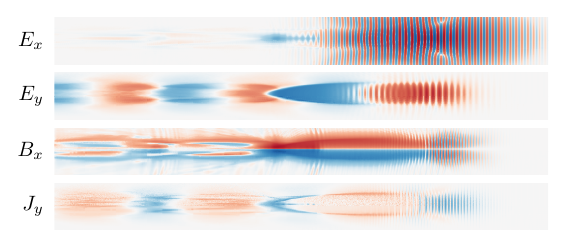}
 \caption{\label{fig:recon}
  Input data (each panel: top) vs. reconstructions (each panel: bottom) for selected components of the electric field ($E_x$, $E_y$), magnetic field ($B_x$) and the current ($J_y$) from a simulation snapshot at $t=\num{57047}$~simulation steps. The $E_x$ plot shows the transverse laser field dominating the picture, while $E_y$ shows the accelerating field. In the longitudinal current $J_y$ an accelerated electron bunch riding the laser wake can be seen as a blue droplet at the center of the panel.}
\end{figure}

The first metric of success is the model's ability to perform autoencoding on held-out data. 
By comparing the original simulation slices with the decoded outputs, we can assess the fidelity of the learned representation. 
As shown in Fig.~\ref{fig:recon}, the model successfully reconstructs the complex structures of the plasma wake and the accelerating cavity in a late time regime at $t=\num{57047}$ simulation steps with very complex dynamics, 
demonstrating that the latent space captures the essential dynamical features of the \gls{lwfa} process. 
Averaged over held-out samples, the masked-encoding reconstruction loss is \num{4.0e-4}, confirming that the encoder recovers the masked field regions consistently.

To analyze the topological structure of this latent space, we employ \gls{pca}, which allows us to project the high-dimensional features into a lower-dimensional manifold while preserving the directions of maximum variance. For this analysis, we aggregate latent tokens from a total of $3\times32\times3$ crops extracted from the $144\times1536\times144$ simulation volume. Each $48^3$ crop is partitioned into $8^3=512$ patches (each of size $6^3$), where each patch is mapped to a 1024-dimensional latent vector. This results in a dataset of $512\times288$ latent vectors of dimension $1024$.

\begin{figure}
 \centering
 \includegraphics[width=\linewidth]{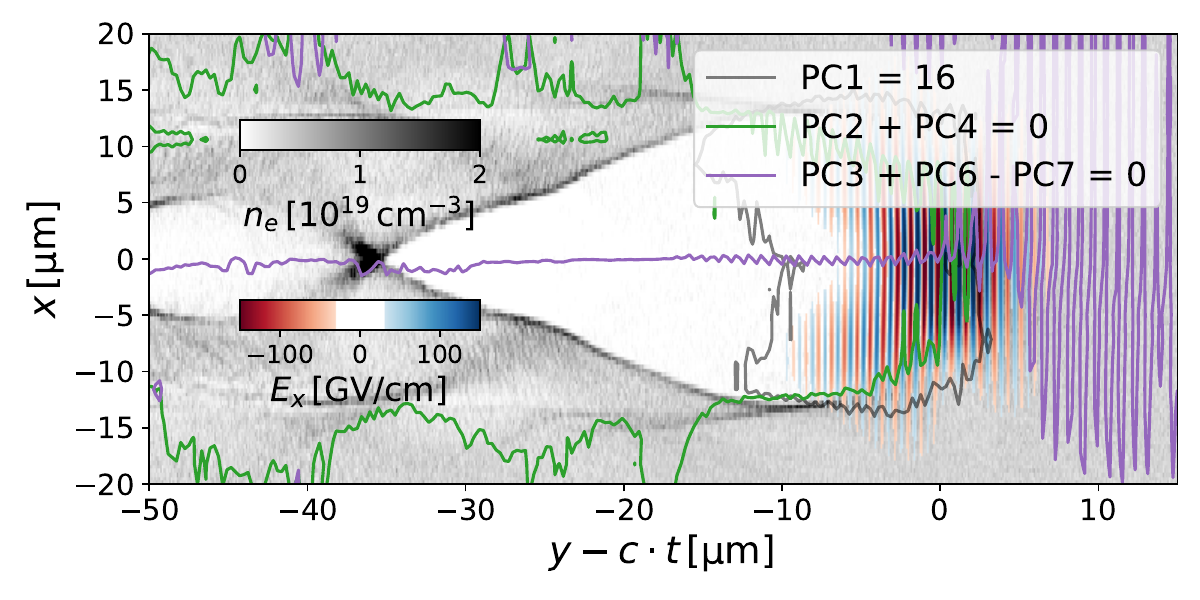}
 \caption{\label{fig:PCA}
 A snapshot of one simulation output depicting the electric field $E_x$ and electron density $n_e$ over the co-moving longitudinal position $y-ct$ and a transverse position $x$. 
 Three contour lines of \gls{pc} linear combinations are overlayed. 
 }
\end{figure}

By projecting the high-dimensional latent tokens onto a 2D map (Fig.~\ref{fig:PCA}), we observe that simulations with similar physical characteristics (e.g., similar acceleration gradients) cluster together. 
This figure shows the electric field component $E_x$ of the linearly polarized laser (included in the training) and the electron density $n_e$ (not included in the training) along the co-moving laser propagation direction $z - c t$ and the transverse position $x$ parallel to the laser polarization. 
The figure also shows the contour lines of certain linear combinations of the individual \glspl{pc} obtained from \gls{pca} of high-dimensional latent tokens. 
We observed that the strongest \gls{pc} (i.e. PC1) usually encloses the laser field (black line), while a certain other combination clearly shows the symmetry axis (purple line), and another the plasma perturbation region (green line).
This indicates that the \gls{mae} objective has naturally organized the data according to the underlying physics of the system, despite the absence of explicit labels during training.

\subsubsection{\gls{lwfa} Spectrum Estimation}

\sectionGuide{1 page}{Richard,Vedhas,Jeffrey}
{
\begin{itemize}
 \item Maybe split into two subsections: Learned represntations, LFWA use-case
  demonstration? -> subsubsections?:
  \begin{enumerate}
  \item autoencoding: show input output pair(s)
  \begin{itemize}
   \item input-decoded pair of full simulation slice. some interesting snapshot, usually $t\approx 80\,000$
   \item show some latent PCA map, ideally with some relation to the I/O pair explained
  \end{itemize}
  \item LWFA spec estimate from expert based on decoding (ideally vs. esimate from input)
  \item LWFA linear probe espec estimate
  \end{enumerate}
\end{itemize}
}

% \todo[inline]{[Placeholder---summary of a planned companion analysis.] There,
% the same target---the mean (peak position) of the Gaussian fit to the
% spectrum---will be predicted directly from the \emph{decoded} accelerating
% field: a domain expert locates the accelerating feature in the input space and
% then relates it to the target through a linear relation. The linear probe
% described below takes a different route: it operates on the \emph{latent} space
% rather than the decoded input space, and it does not take the exact location of
% the feature---the selected crops only roughly narrow it down.}

\begin{figure}
 \centering
 \includegraphics[width=\linewidth]{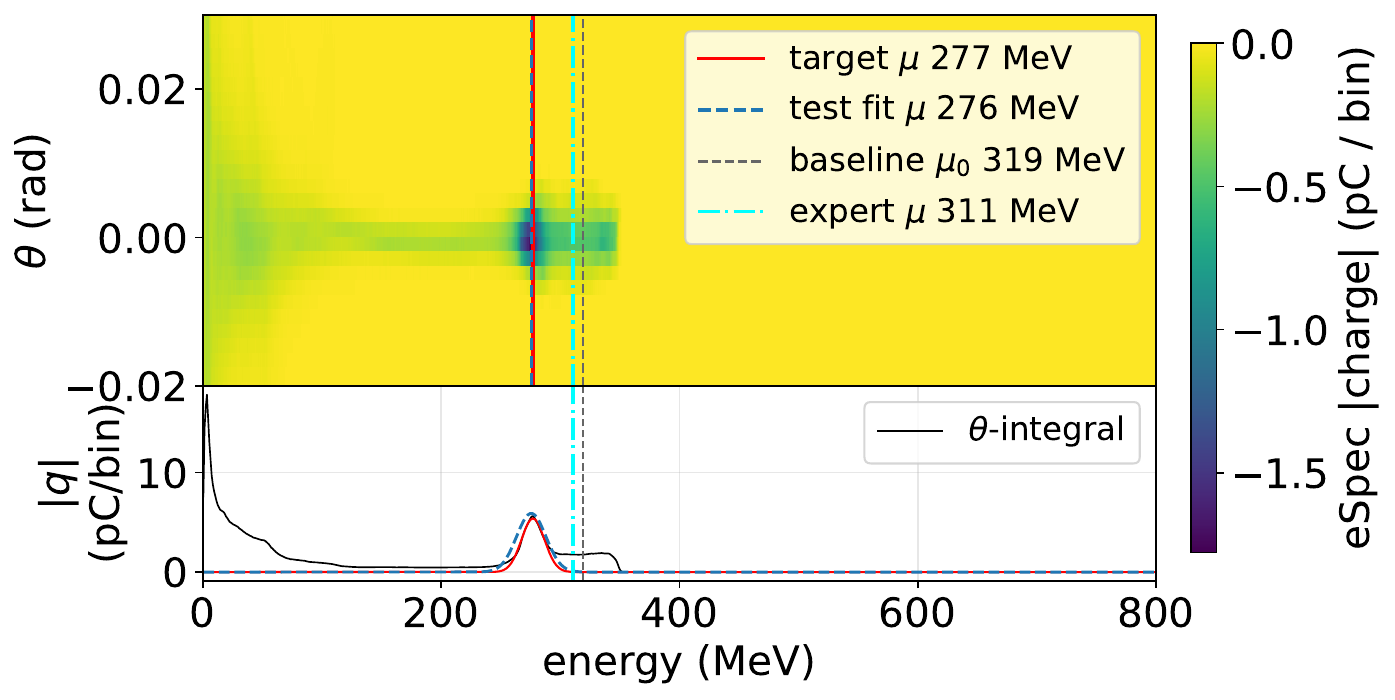}
 \caption{\label{fig:espec}
 Electron energy spectrum (integrated over the full $\theta$ range) of a
 held-out simulation. The model-predicted Gaussian peak shown is the mean over the
 seven evaluated crops, at propagation-axis offsets $y=672$--$960$ cells; it
 gives $\mu \approx \qty{353}{\MeV}$, close to the target peak of
 $\approx \qty{365}{\MeV}$ and well above the training-mean baseline of
 $\approx \qty{319}{\MeV}$. An expert prediction based on the accelerating field close to the time of electron bunch self-injection, calculated from the original simulation output, is shown for comparison.}
\end{figure}

The electron energy spectrum is a critical observable of \gls{lwfa}. 
To test whether the self-supervised representation learned by the
Recycler encodes physically relevant information in a way useful for downstream
models, we freeze the trained encoder and fit a
\emph{linear} probe from its latent tokens to a single target:
the mean energy of the accelerated electron bunch, defined as the mean $\mu$ of
a Gaussian fit to the hind part of the electron spectrum (see Fig.~\ref{fig:espec}).
While the mean energy is approximately linearly related
to the accelerating electric field inside the accelerating cavity, extracting
the field from raw simulation data is less straightforward algorithmically than
it is for a domain expert.
% Unlike the
% decoded-field analysis, the probe does not map back to the input space and does
% not take the exact location of the accelerating feature---the selected crops
% only roughly narrow down where it forms.
The linear probe also tests whether the learned representation exposes features
related to the accelerating field in a linear way. 
For \gls{mae} pre-training to be successful, i.e. for the decoder to learn to
predict the fields and currents in masked patches, the encoder must learn to
extract features from available patches which predict the masked regions. 
While there is no explicit physics-based loss function, relations found by the model
may resemble physical laws which relate fields and currents anywhere in the
input volume to the substantive linearly accelerating electric field. 
This information must be encoded in the latent in a way which is accessible enough for
the shallow decoder to succeed at reconstructing the masked patches, and thus
may be accessible enough for a linear probe with a simpler target.

As the \gls{mae} works on crops of $48^3$ cells, which are ingested as 512 tokens, we
focus the probe on crops which are most likely to encode the relevant feature
directly, i.e. transversely centered crops along the laser propagation axis.
First we use a \gls{pca} fitted on the entire \gls{mae} training data to reduce the token
dimension, from the 512 decoder dimensions to $k=60$.
To further reduce the input dimension of the problem, we only input a random
subset of ten encoded tokens, less than the \qty{10}{\percent} used in \gls{mae} pre-training, into
the linear model, so the fit
rests on a small, sparsely sampled part of the latent description. We limit the input to the model to two snapshots around the time of injection, which can be estimated from a zero-crossing in an early particle-energy histogram, with the addition of the simulation time of the \emph{first} snapshot of the pair. All information going into the model, including the estimation of injection time, is derived only from simulation output available before $t=\num{50000}$~simulation time steps, for a simulation with a total length of \num{110000} steps.

The probe is
trained and evaluated with a \emph{run-level} split of the \gls{mae}'s training set:
every crop of a simulation
is used entirely for training or entirely for validation, and no run appears in
both, so the reported numbers are for simulations unseen by the linear model.
The described collection of features yields a linear model with \num{1202} parameters (including bias), far exceeding the $n=77$ simulations available for offline training. Hence, we expand the dataset by using each crop as a separate training sample and further augment by selecting ten random input tokens 63 times. 

\begin{figure}
 \centering
 \includegraphics[width=\linewidth]{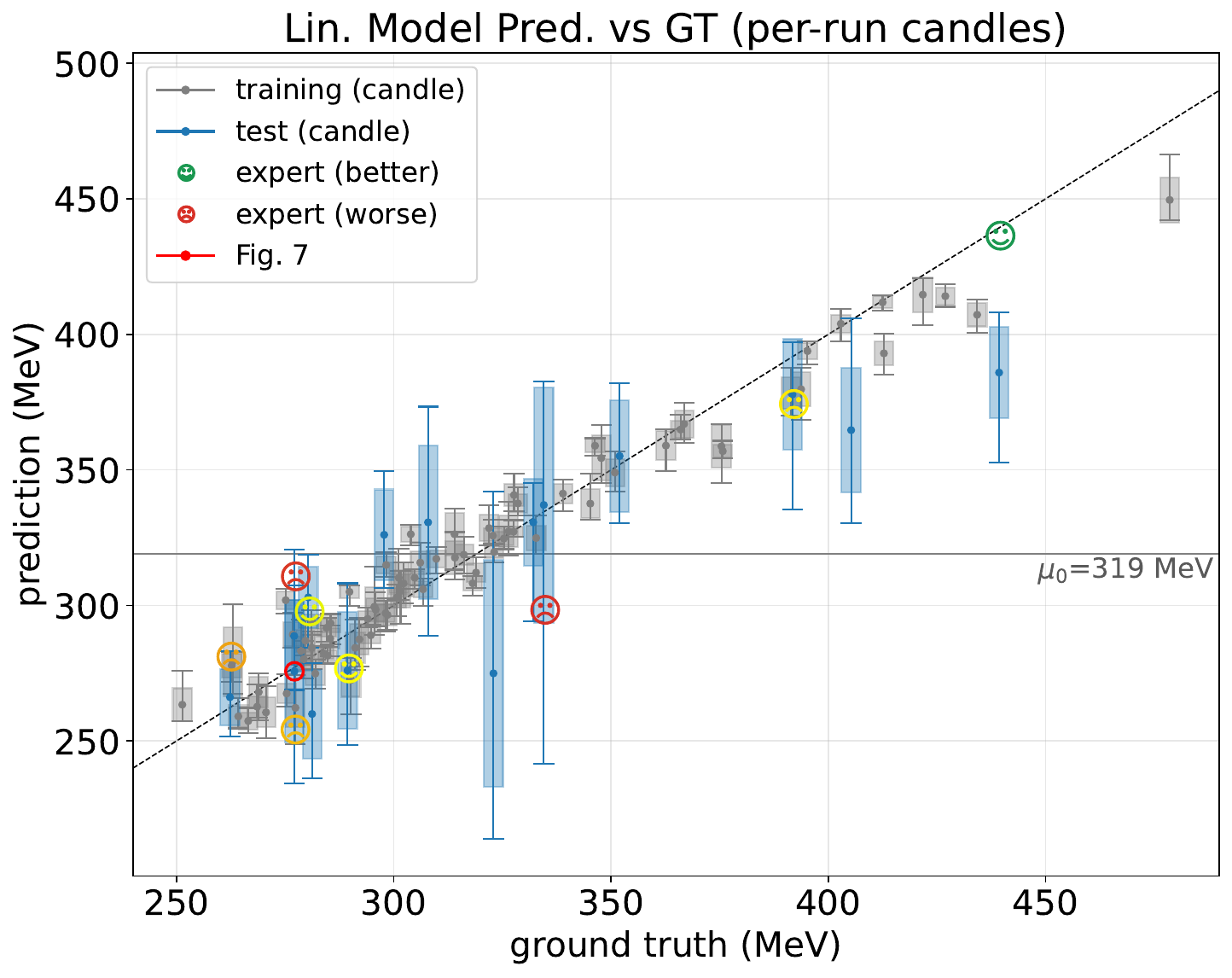}
 \caption{\label{fig:especscatter}
 Predicted mean electron bunch energy vs. ground-truth. The candles summarize linear model predictions from the
 seven evaluated crops, at propagation-axis offsets $y=672$--$960$ cells, for each simulation. The mean across crop predictions is evaluated against the expert prediction.}
\end{figure}

On held-out runs $\mu$ is recovered with
$R^2 \approx 0.59$ (Pearson $r \approx 0.79$) averaged over crops. When crops closer to the left border of the simulation
box are disregarded during \emph{evaluation}, this improves to $R^2 \approx
0.59$--$0.76$, which demonstrates that the model itself can show which regions
are good predictors and pre-selection of crops is not required. Fig.~\ref{fig:especscatter} shows the predictions on the training data ($n=77$ simulations) and on held-out test data. Hyper parameters were optimized on a validation set of another $n=22$ simulations, which are not shown in the figure.
% restricted to the plasma region where the acceleration occurs after the laser focus.
The comparison of training and test performance shows that the linear model clearly generalizes across the domain covered by our simulations.
Control fits confirm that the
signal is physical rather than an artifact: fitting the simulation iteration
alone carries no predictive power ($R^2 \approx 0$), a random linear \gls{pca} direction
yields $R^2 \approx 0.05$, while a single leading \gls{pca} component already reaches
$R^2 \approx 0.3$. The in-transit-trained representation thus contains a
linearly accessible encoding of the electron energy spectrum, demonstrating
that it captures the plasma dynamics that drive the acceleration and not
merely the simulation setup.

While the linear model relies on the encoder encoding relevant accelerating-field information in fewer dimensions than make up the raw simulation data, a laser plasma physicist (domain expert) can also identify the relevant values in the raw simulation data during manual analysis. Hence, to further evaluate the efficacy of the learned representations, we benchmark the linear-model predictions against domain-expert estimations.

For the ``expert'' estimation, the on-axis peak current $y_\mathrm{inj} = \mathrm{argmax}(J_y)$ is used to determine the accelerating electric field $E_y(y_\mathrm{inj})$. 
For each of the 11 snapshots stored on disk at random intervals during each simulation, the gained energy is assumed to increase under a similar electric field until the next snapshot
\begin{equation}
    E_\mathrm{peak} = -q_e \int\limits_{y_\mathrm{start}}^{y_\mathrm{end}} E_y(y) \mathrm{d}y \approx -q_e \cdot c \sum \limits_{t_\mathrm{start}}^{t_\mathrm{end}} E_y(y_\mathrm{inj}) \cdot \Delta T \label{eq:espec}
\end{equation}
%
%\atrp{how many snapshots does the expert prediction take into account? I thought the filed is just assumed to be constant from the time of injection to the end, the integral makes the estimation more involved than I thought it would be.}
%
with $q_e$ denoting the electron charge, $c$ the speed of light, $\Delta T$ the time duration between snapshots and $E_\mathrm{peak}$ the estimate for the final peak energy. 

%\todo{This still needs to be better integrated with the rest of the subsection -- like so?}
Fig.~\ref{fig:especscatter} shows that estimating the peak energy from the raw data snapshots using \eqref{eq:espec} leads to similar results as the linear model, confirming that the \gls{vit} encodes enough information into its latent to linearly predict a relevant physical simulation outcome to comparable accuracy as a linear approximation grounded in physics. 
Furthermore, the model is able to make this prediction using only information before the half-way point of each simulation run ($t\lesssim \num{50000}$ simulation steps) and based on only 77 training simulations. 
In contrast, the ``expert'' estimation for each simulation requires all snapshots, including late times, and its accuracy depends heavily on the sparsity of the sampling. 
For example, if snapshots only begin after injection, the ``expert'' underestimates the peak energy despite having access to more data. Alternatively, if the time between snapshots, $\Delta T$, is large, the contribution from a single snapshot becomes dominant, thus increasing the variance of the prediction.
If the domain ``expert'' were only to use the same data the linear model uses, its prediction would be even worse. 

%\atrp{can we make this claim in comparison to the expert prediction? - Done}

%\section{Discussion}

\sectionGuide{1 page}{Jeffrey,Richard}
{
\begin{itemize}
 \item ~
\end{itemize}
}

\section{Conclusion}

\sectionGuide{1 page}{Micha,Jeffrey,Richard,Sunita}
{
\begin{itemize}
 \item ~
\end{itemize}
}

We have demonstrated for the first time that it is possible to train a large-scale model in-transit with data streaming from multiple simulations running concurrently on an exascale high performance system. We have intentionally chosen the application case of highly nonlinear laser wakefield acceleration (LWFA) of electrons in a plasma, where high-fidelity data is key to predicting operational parameters such as the electron beam energy crucial for applications of LWFA. We have also shown that this model can predict these parameters, an important result towards our goal of replacing long-term storage of high-fidelity data by training a \emph{frontier model}.
Our current scaling results highlight the difficulty of running such workflows efficiently up to the full scale of machines such as Frontier. This is an important finding as it illustrates the challenge of solving a complex optimization problem that includes optimizing single simulations, the orchestration of multiple simulations, data locality and throughput of streaming and the parallelization of model training. Our preliminary analysis suggests that scaling efficiency can be increased by adding further dimensions of parallel processing to increase model throughput without increasing the size of DDP's global all-reduce, such as model pipelining and tensor sharding and also training ensembles of models which enable uncertainty estimation.
Although efficient scaling must take into account the specifics of the simulations and the model, the overall approach we take can be easily adapted to other simulation codes, applications and challenges. Consequently, our next goal is to intelligently steer simulation campaigns towards finding optimum operational parameters. On top of that, we aim to query the model, via drop-out or as an ensemble, to decide if new simulation data is needed or if the model can already predict the outcome of a certain simulation, thus over time reducing the amount of new simulations needed through an active-learning loop~\cite{CohnAtlasLadner_1994ImprovinGeneraliActiveLearning}.
With these improvements in place, we will be set to realize in-transit training of large-scale frontier models with data from intelligently steered simulation campaigns at high parallel efficiency. By closing the feedback loop with active learning we can continuously improve model quality while efficiently allocating new simulations. This addresses the central question of continuous learning for frontier models by implementing feedback loops between data production, training and model inference at the exascale.
%
%
\iffalse
\jk{\textit{the following is redundant now: already covered by the previous, riight?}}
This paves the way towards direct steering of simulation campaigns to efficiently covering large parameter spaces via a self-improving feedback loop for creating training data that minimizes the amount of computing time needed.

%The Recycler workflow solved the scheduling problem of obtaining high-fidelity simulation data from multiple simulations for machine learning (ML) training without exceeding the file system capacity. 
%However, we had to write our own scheduling procedure for the demonstrator due to limitations on communication between independent SLURM jobs, commonly imposed by HPC systems such as Frontier. 
%If HPC systems would allow fast communication between jobs, it would ease the deployment of this streaming method and enable a much larger community to use it. 
goals, add limitations (scaling efficiency ), ways to address 
\begin{itemize}
    \item use full system
    \item `steer and close the loop' (obstacles: what sim to look at,)
    \item vast amounts of sim campaigns, huge corpora of knowledge system
\end{itemize}
\fi

\section*{Acknowledgments}
\footnotesize{
{
%\begin{itemize}
% \item please check/add/remove funding and compute time acknowledgments
%\end{itemize}
%}

This research was supported 
by Laserlab-Europe EU-H2020 GA no. 871124 and the European Union through Grant Agreement No. 101093261 (Plasma-PEPSC). 
This research used resources of the Oak Ridge Leadership Computing Facility at the Oak Ridge National Laboratory, supported by the Office of Science of the U.S. Department of Energy under Contract No. DE-AC05-00OR22725. This work was supported in part by the OLCF Center for Accelerated Application Readiness (CAAR) program. The support work from the ADIOS team was supported by the U.S. Department of Energy, Office of Science, Office of Advanced Scientific Computing Research, Scientific Discovery through Advanced Computing (SciDAC) program, under the “RAPIDS Institute”.

This research was supported by the Helmholtz Association Initiative and
Networking Fund in the frame of Helmholtz AI and the project NEOPIC
(ZT-I-PF-5-229). This work was supported by Helmholtz AI computing resources (HAICORE) of the Helmholtz Association’s Initiative and Networking Fund through Helmholtz AI at at Forschungszentrum Jülich.
%\end{itemize}
}
}

\bibliography{bibliography}
\bibliographystyle{ieeetr}

\end{document}